\documentclass[twocolumn,showkeys,aps,prb,showpacs]{revtex4-1}
\usepackage{graphicx}
\usepackage[CJKbookmarks,dvipdfm,colorlinks,linkcolor=blue,citecolor=blue]{hyperref}

\begin{document}

\title{Elastic and transport properties of topological semimetal ZrTe}

\author{San-Dong Guo,  Yue-Hua Wang and Wan-Li Lu}
\affiliation{School of Physics, China University of Mining and
Technology, Xuzhou 221116, Jiangsu, China}

\begin{abstract}
Topological semimetal may have substantial applications in electronics,
spintronics and quantum computation.  Recently, ZrTe is predicted as  a new type of topological semimetal due to coexistence of
Weyl fermion and massless triply degenerate nodal points. In this work, the elastic and transport properties of ZrTe are investigated by  combining the first-principles calculations and semiclassical Boltzmann transport theory. Calculated  elastic  constants prove mechanical stability of ZrTe, and the bulk modulus, shear modulus, Young's modulus  and Possion's ratio also are calculated. It is found that spin-orbit coupling (SOC)  has slightly enhanced effects on
Seebeck coefficient, which  along a(b) and c directions for
pristine ZrTe  at 300 K is  46.26 $\mu$V/K and 80.20  $\mu$V/K, respectively.
By comparing the experimental  electrical conductivity of ZrTe (300 K)  with calculated value, the scattering time is determined for 1.59 $\times$ $10^{-14}$ s. The predicted room-temperature electronic thermal conductivity along a(b) and c directions is 2.37 $\mathrm{W m^{-1} K^{-1}}$  and  2.90 $\mathrm{W m^{-1} K^{-1}}$, respectively. The  room-temperature lattice thermal conductivity  is predicted as  17.56   $\mathrm{W m^{-1} K^{-1}}$ and  43.08   $\mathrm{W m^{-1} K^{-1}}$ along a(b) and c directions, showing very strong anisotropy.
Calculated results show that isotope scattering produces observable effect on lattice thermal conductivity.  To observably reduce lattice thermal  conductivity by nanostructures,  the  characteristic length should be smaller than 70 nm, based on  cumulative lattice thermal conductivity with respect to phonon mean free path(MFP) at 300 K.
It is noted that average room-temperature lattice thermal conductivity of ZrTe is  slightly higher than that of isostructural MoP, which is due to larger   phonon lifetimes and smaller Gr$\mathrm{\ddot{u}}$neisen parameters. Finally, the total thermal conductivity as a function of temperature is predicted for pristine ZrTe. Our works provide  valuable informations for ZrTe-based nano-electronics devices, and motivate  further experimental works to study  elastic and transport properties of ZrTe.

\end{abstract}

\pacs{72.15.Jf, 71.20.-b, 71.70.Ej, 79.10.-n ~~~~~~~~~~~~~~~~~~~~~~~~~~~~~~~~~~~Email:guosd@cumt.edu.cn}

\maketitle

\section{Introduction}
 From topological insulator to semimetal, the recent discovery of new type of topological nontrivial phase
has sparked intense research interest in condensed matter physics and material science\cite{q6,q7,q8,q1,q2,q3,q4,q10,n1,q5,q5-1}.
The representative topological semimetals include Dirac semimetals, Weyl semimetals and nodal line semimetals\cite{q4,q5,q8},
such as $\mathrm{Na_3Bi}$ as a classic  Dirac Semimetal\cite{q4}, TaAs as a  representative Weyl semimetal\cite{q10} and ZrSiS as a  typical nodal line semimetal\cite{n1},  which have
been confirmed by angle-resolved photoemission spectroscopy (ARPES).
In Dirac   and Weyl semimetals, four-fold degenerate Dirac point and
two-fold degenerate Weyl point can be observed in the momentum space\cite{q4,q10,q5,q5-1}, while two bands cross in the form of a periodically
continuous line or closed ring for node-line semimetals\cite{q8,n1}.

Beyond Dirac and Weyl fermions, some new types of topological semimetals are proposed, which are  identified by
three-, six- or eight-fold band crossings\cite{j1}. By the crossing of a double-degeneracy band and
a non-degeneracy band,  the three-fold degenerate crossing points are predicted  in the materials with WC-type structure,
such as MoP, WC and ZrTe\cite{q11,q11-0}, and in $\mathrm{InAs_{0.5}Sb_{0.5}}$\cite{q11+0}.
Then, the ARPES declares  the presence of a triply degenerate point in  MoP, and pairs of Weyl points  coexist with the three-component fermions\cite{q7}.
Experimentally, the highly metallic characteristics with remarkably low resistivity  and high mobility (2 K) has been found in  MoP\cite{q13-1}.
The recent ARPES experiments and transport measurements have also
shown that WC has  the nontrivial topological nature\cite{j2,j3}. The experimental results of the magnetoresistance, Hall effect, and quantum
Shubnikov-de Haas oscillations on single crystals of ZrTe have been reported, indicating  ZrTe with low carrier density, high carrier mobility, small cross-sectional area of Fermi surface, and light cyclotron effective mass\cite{j4}.

\begin{figure}
  \includegraphics[width=8cm]{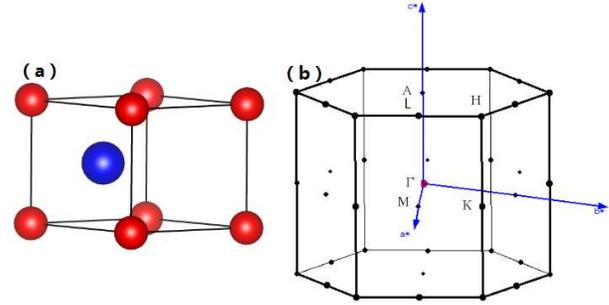}
  \caption{(Color online) (a)The crystal structure of ZrTe in one unit cell, and the blue and red balls represent Zr and Te atoms. (b)the  Brillouin zone with high-symmetry points.}\label{st}
\end{figure}

Recently, the elastic and thermal transport properties of some representative topological semimetals have been investigated, such TaAs and MoP\cite{q12,q13,j5,j6}.
The lattice thermal conductivity of both TaAs and MoP  shows  obvious anisotropy
along the a(b) and c crystal axis\cite{q12,q13,j6}. High thermoelectric performance of TaAs has  been predicted, and the maximum thermoelectric figure of merit $ZT$ is  up to 0.63 (900 K) in n-type doping along  c direction\cite{q13}.
In this work, the elastic and  transport properties of three-fold degeneracy topological semimetal ZrTe are investigated by  combining the first-principles calculations and semiclassical Boltzmann transport theory. The elastic  constants,  bulk modulus, shear modulus, Young's modulus  and Possion's ratio  are predicted with the generalized gradient approximation (GGA). The  electronic  transport coefficients are also calculated using both GGA and GGA+SOC.
It is found that SOC  has slight influences on electronic  transport coefficients. For
pristine ZrTe, the Seebeck coefficient,  electrical conductivity,  power factor  and   electronic thermal conductivity are calculated, which can be verified by future experiments.
The lattice thermal conductivity as a function of temperature is predicted within GGA, which shows a distinct anisotropic  property along the a(b) and c crystal axis.
Similar results can be found in  topological semimetals TaAs and MoP\cite{q12,q13,j6}. The isotope and size effects on the lattice thermal conductivity are also studied, and the phonon mode analysis is also  performed to
understand deeply phonon transport of ZrTe. The total lattice thermal conductivity ($\kappa$=$\kappa_L$+$\kappa_e$) as a function of temperature
is also predicted for pristine ZrTe.
This work sheds  light on the elastic and  transport properties of ZrTe, and could offer valuable guidance for MoP-based nano-electronics devices.
\begin{figure}
  \includegraphics[width=8cm]{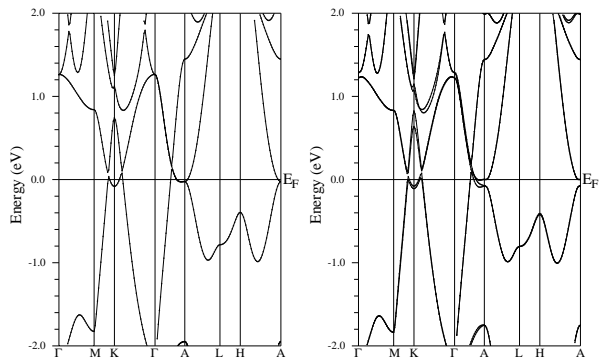}
  \caption{The calculated energy  band structures of ZrTe along high-symmetry paths with GGA (Left) and GGA+SOC (Right).}\label{band}
\end{figure}

The rest of the paper is organized as follows. In the next
section, we shall give our computational details. In the third section, we shall present elastic and  transport properties of ZrTe. Finally, we shall give our conclusions in the fourth section.

\begin{figure*}[!htb]
  \includegraphics[width=12cm]{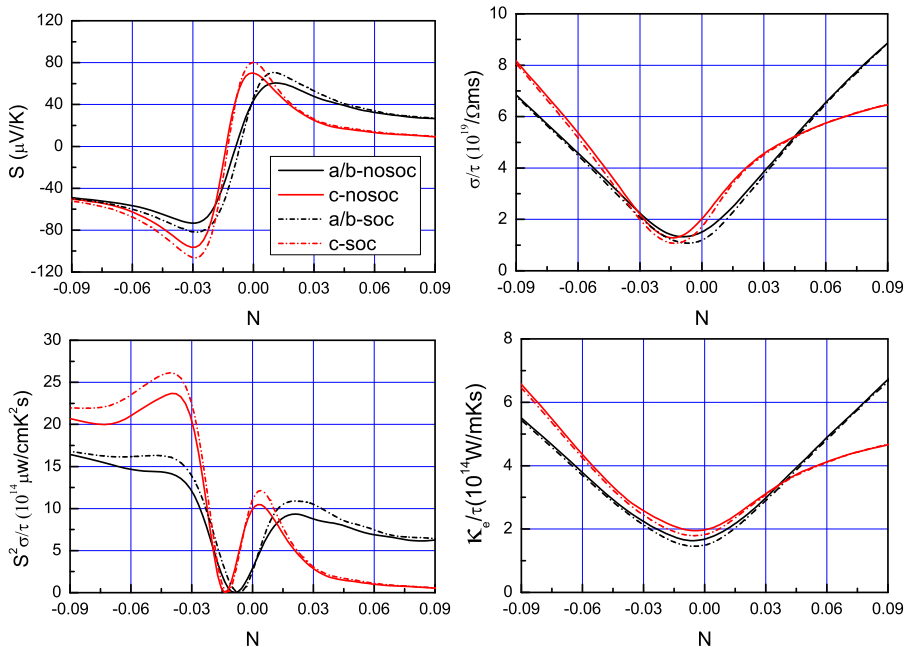}
  \caption{(Color online) At room temperature (300 K),   Seebeck coefficient S,  electrical conductivity with respect to scattering time  $\mathrm{\sigma/\tau}$, power factor with respect to scattering time $\mathrm{S^2\sigma/\tau}$  and  electronic thermal conductivity with respect to scattering time $\mathrm{\kappa_e/\tau}$ of ZrTe  as a function of doping level (N) with GGA and  GGA+SOC.}\label{s}
\end{figure*}

\section{Computational detail}
Within the density functional theory (DFT)\cite{1}, a full-potential linearized augmented-plane-waves method, using   GGA of Perdew, Burke and  Ernzerhof  (GGA-PBE)\cite{pbe}, is employed to investigate electronic structures of  ZrTe, as implemented in the  WIEN2k code \cite{2}.
 The SOC is included self-consistently \cite{10,11,12,so}, which produces observable effects on electronic transport coefficients. The convergence results are determined by using  4000 k-points in the
first Brillouin zone (BZ) for the self-consistent calculation, making harmonic expansion up to $\mathrm{l_{max} =10}$ in each of the atomic spheres, and setting $\mathrm{R_{mt}*k_{max} = 8}$ for the plane-wave cut-off. The self-consistent calculations are
considered to be converged when the integration of the absolute
charge-density difference between the input and output electron
density is less than $0.0001|e|$ per formula unit, where $e$ is
the electron charge.
Based on calculated energy band
structures, transport coefficients of electron part are calculated through solving Boltzmann
transport equations within the constant
scattering time approximation (CSTA),  as implemented in
BoltzTrap code\cite{b}. To obtain accurate transport coefficients,
 the parameter LPFAC  is set as 10, and  2772 k-points is used in the  irreducible BZ for the calculations of energy band structures.
\begin{table}[!htb]
\centering \caption{The experimental lattice constants $a$ and $c$ ($\mathrm{{\AA}}$); elastic constants $C_{ij}$, bulk ($B$), shear ($G$) and Young's ($E_{xx}$ and $E_{zz}$) moduli  (in GPa); Poisson's ratio ($\nu$). }\label{tab}
  \begin{tabular*}{0.48\textwidth}{@{\extracolsep{\fill}}ccccccc}
  \hline\hline
$a$ & $c$ & $C_{11}$ & $C_{12}$& $C_{13}$&$C_{33}$ &$C_{44}$\\\hline\hline
3.771&3.861&140.82&58.78&88.81&201.11  &110.36\\\hline
$B$&$G$&$E_{xx}$&$E_{zz}$&$\nu_{xy/yx}$& $\nu_{xz/yz}$ &$\nu_{zx/zy}$\\\hline
102.50&61.26&97.83&122.07&0.19& 0.36 &0.45\\\hline\hline
\end{tabular*}
\end{table}

For elastic properties and phonon transport, the first-principles calculations are performed within the projected augmented wave (PAW) method, and the GGA-PBE is adopted as exchange-correlation energy functional,
as implemented in the VASP code\cite{pv1,pv2,pbe,pv3}.
A plane-wave basis set is employed with
kinetic energy cutoff of 400 eV,  and the electronic stopping criterion is $10^{-8}$ eV.
The  lattice thermal conductivity of  ZrTe is performed by solving linearized phonon Boltzmann equation with the single mode relaxation time approximation (RTA),   as implemented in the Phono3py code\cite{pv4}. The lattice thermal conductivity can be expressed as
\begin{equation}\label{eq0}
    \kappa_L=\frac{1}{NV_0}\sum_\lambda \kappa_\lambda=\frac{1}{NV_0}\sum_\lambda C_\lambda \nu_\lambda \otimes \nu_\lambda \tau_\lambda
\end{equation}
where $\lambda$ is phonon mode, $N$ is the total number of q points sampling the BZ, $V_0$ is the volume of a unit cell, and  $C_\lambda$,  $ \nu_\lambda$, $\tau_\lambda$   is the specific heat,  phonon velocity,  phonon lifetime.
The interatomic force constants (IFCs) are calculated by
the finite displacement method.
 The second-order harmonic IFCs
are calculated using a 4 $\times$ 4 $\times$ 4  supercell  containing
128 atoms with k-point meshes of 2 $\times$ 2 $\times$ 2.  Using the harmonic IFCs, phonon dispersion of ZrTe can be attained by  Phonopy package\cite{pv5}.
 The  group velocity  and specific heat can be attained from phonon dispersion
which also  determines the allowed three-phonon scattering processes. The third-order anharmonic IFCs are calculated using a 3 $\times$ 3 $\times$ 3
supercells containing 54 atoms with k-point meshes of 3 $\times$ 3 $\times$ 3, and  the total number of displacements
is 508. Based on third-order anharmonic IFCs,  the phonon lifetimes can be attained from the three-phonon scattering. To compute lattice thermal conductivities, the reciprocal spaces of the primitive cells  are sampled using the 20 $\times$ 20 $\times$ 20 meshes.

\begin{figure*}[!htb]
  \includegraphics[width=16cm]{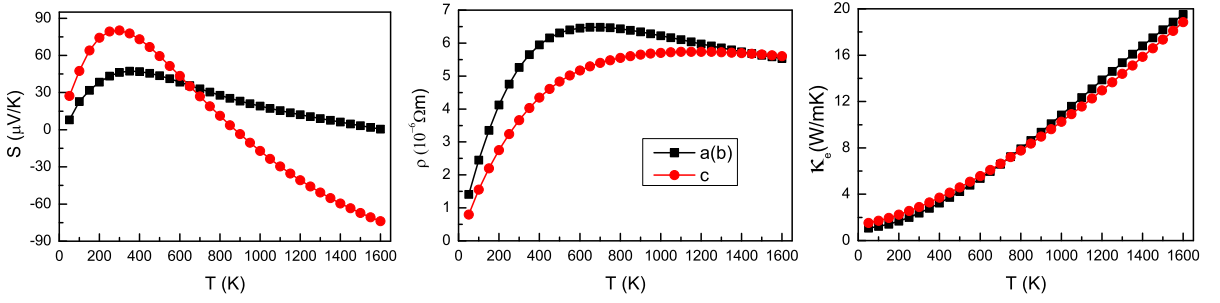}
  \caption{(Color online)For  pristine ZrTe, the  Seebeck coefficient S,  electrical resistivity $\rho$ and  electronic thermal conductivity $\mathrm{\kappa_e}$  as a function of temperature with GGA+SOC.}\label{s-t}
\end{figure*}

\begin{figure}
  \includegraphics[width=8cm]{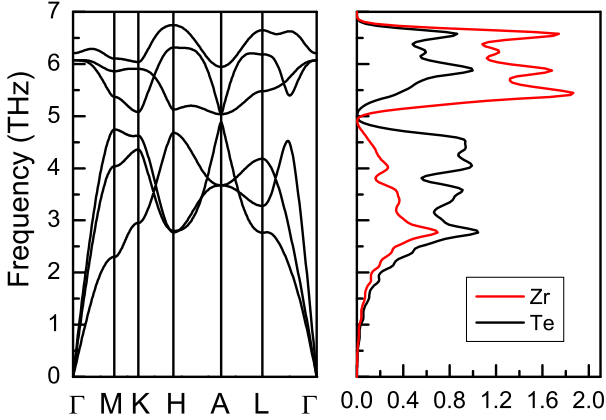}
  \caption{(Color online)Phonon dispersions of ZrTe with the corresponding  atom partial DOS (PDOS). }\label{ph}
\end{figure}

\section{MAIN CALCULATED RESULTS AND ANALYSIS}
\subsection{ELASTIC PROPERTIES}
 The WC-type ZrTe  possesses  space group  $P\bar{6}m2$ (No. 187),  with Zr and
Te atoms occupying the 1d (1/3, 2/3, 1/2) and 1a (0,0,0) Wyckoff positions, respectively.
 The  crystal structure and BZ of ZrTe are shown  in \autoref{st}.
All the results attained in the following are from the calculations with experimental lattice constants ($a$=$b$= 3.7707 $\mathrm{{\AA}}$, $c$=3.8606 $\mathrm{{\AA}}$ )\cite{q14}.  The most basic  physical quantities of elastic properties are  elastic constants $C_{ij}$, which can be used to construct else elastic physical quantities. The elastic constants are a four-rank tensor. However, the elastic constants
are reduced to five independent ones: $C_{11}$, $C_{12}$, $C_{13}$, $C_{33}$, $C_{44}$ due
to the symmetry for hexagonal crystal, and $C_{66}$ can be obtained by $(C_{11}-C_{12})/2$. The calculated $C_{ij}$ are
shown in \autoref{tab}. To prove mechanical stability of ZrTe, we use the following mechanical
stability criterion for the  hexagonal materials\cite{el,q15}:
\begin{equation}\label{e1}
C_{44}>0
\end{equation}
\begin{equation}\label{e1}
 C_{11}>|C_{12}|
\end{equation}
\begin{equation}\label{e1}
(C_{11}+2C_{12})C_{33}>2C_{13}^2
\end{equation}
By simple calculations,   these criteria are satisfied for ZrTe,
which means no strong tendency to become unstable with the increasing pressure.

The Voigt's, Reuss's and Hill's bulk modulus can be attained by the following equations:
\begin{equation}\label{4}
    B_V=\frac{1}{9}(2C_{11}+C_{33}+2C_{12}+4C_{13})
\end{equation}
\begin{equation}\label{5}
    B_R=(2S_{11}+S_{33}+2S_{12}+4S_{13})^{-1}
\end{equation}
\begin{equation}\label{6}
    B_H=\frac{1}{2}(B_V+B_R)
\end{equation}
The Voigt's, Reuss's and Hill's shear modulus can be calculated by using these formulas:
\begin{equation}\label{4}
    G_V=\frac{1}{15}(2C_{11}+C_{33}-C_{12}-2C_{13}+6C_{44}+3C_{66})
\end{equation}
\begin{equation}\label{5}
    G_R=[\frac{1}{15}(8S_{11}+4S_{33}-4S_{12}-8S_{13}+6S_{44}+3S_{66})]^{-1}
\end{equation}
\begin{equation}\label{6}
    G_H=\frac{1}{2}(G_V+G_R)
\end{equation}
The $S_{ij}$'s can be obtained by inverting the elastic constants matrix. The calculated Voigt's, Reuss's and Hill's bulk modulus   are 106.17 GPa,   98.82 GPa and 102.50 GPa, respectively, and 68.77 GPa,  53.75 GPa and 61.26  GPa for shear modulus. The Hill's bulk and shear modulus are listed in \autoref{tab}.
The $B$ and $G$ can be used to measure material behaviour as ductile or brittle. If the $B/G$ ratio value is larger than 1.75, the material behaves as ductile, otherwise, it shows a brittle character. For ZrTe, the calculated $B/G$ ratio value with Hill's bulk and shear modulus is 1.67, indicating that the brittle character is dominant. This is different from MoP, where the ductile character is dominant\cite{j6}.

The Young's modulus  $E_{ii}$  can be calculated by the relationship:
\begin{equation}\label{23}
    E_{ii}=1/S_{ii}
\end{equation}
The numerical calculated values  are $E_{xx}$=$E_{yy}$=97.83 GPa and
$E_{zz}$=122.07 GPa, respectively. The Poisson's ratios $\nu_{ij}$ can be calculated by:
\begin{equation}\label{24}
    \nu_{ij}=-S_{ij}/S_{ii}
\end{equation}
 The calculated results are $\nu_{xy}$=$\nu_{yx}$=0.19,
$\nu_{xz}$=$\nu_{yz}$=0.36 and $\nu_{zx}$=$\nu_{zy}$=0.45.
The characteristics of chemical bonds can be reflected by the Poisson's ratio.
For covalent materials, the $\nu$  is low  with a typical value of  0.10, while for ionic materials  with a high typical value of  0.25\cite{n2}. Based on $\nu$ values of ZrTe, ionic bonding is dominant.

\subsection{ELECTRONIC TRANSPORT}
The energy band structures of ZrTe along high-symmetry paths are shown in \autoref{band} using both GGA and GGA+SOC.
Our calculated results agree well with previous theoretical ones\cite{q11}.  It is found that the six-fold degenerated nodal point  splits
 into the two triply degenerate nodal points (TDNPs)  along the $\Gamma$-A direction, when the SOC is included\cite{q11}. However,  the  six pairs of Weyl nodes appear around K point in its first BZ\cite{q11}. Based on calculated energy band structures, the  electronic transport coefficients of ZrTe can be attained  using CSTA Boltzmann theory.   Although the calculated electrical conductivity depends  on  scattering time $\mathrm{\tau}$  using CSTA,  the Seebeck coefficient  is independent of scattering time,  directly compared with  experimental results.
 At room temperature, the Seebeck coefficient S,  electrical conductivity with respect to scattering time  $\mathrm{\sigma/\tau}$,  power factor with respect to scattering time $\mathrm{S^2\sigma/\tau}$  and   electronic thermal conductivity with respect to scattering time $\mathrm{\kappa_e/\tau}$ along a(b) and c directions as a function of doping level using both GGA and GGA+SOC are plotted in \autoref{s}.
 Within the framework of  rigid band approach,  the n- or p-type doping  can be simulated by simply shifting  Fermi level into conduction  or valence bands,    which  is effective in low doping level\cite{q30,q31,q32}.
 \begin{figure}
  \includegraphics[width=8cm]{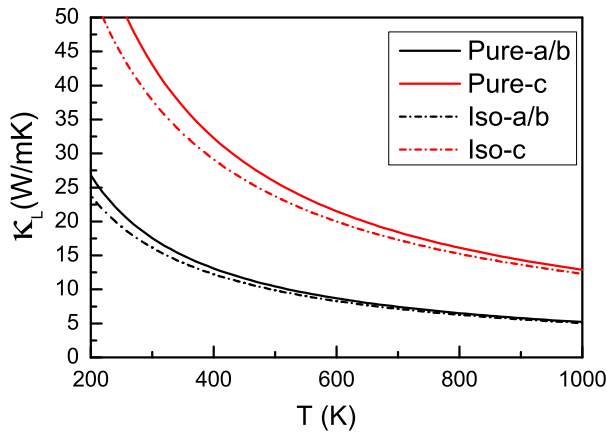}
  \caption{(Color online) The lattice thermal conductivities  of pure and isotopic ZrTe  as a function of temperature, including a(b)  and c directions. }\label{kl}
\end{figure}

Calculated results show that SOC has a slightly enhanced  effect on n- and p-type  Seebeck coefficient (absolute value) along both a(b) and c directions in low doping level. It is found that the  Seebeck coefficients are relatively strongly anisotropic along a(b) and c directions.
The calculated S along a(b) and c directions for
pristine ZrTe is 46.26 $\mu$V/K and 80.20  $\mu$V/K, respectively,  which can be  farther verified by the experiment.
When the Seebeck coefficient vanishes, the doping level along a(b) and c directions  is about -0.0065 and -0.0124, respectively.
It is found that n-type doping has more excellent  Seebeck coefficient than p-type doping.
For $\mathrm{\sigma/\tau}$, the slightly reduced influence caused by SOC can be observed along both a(b) and c directions.
The electronic thermal conductivity  $\mathrm{\kappa_e}$ relates to the electrical
conductivity  $\mathrm{\sigma}$ via the Wiedemann-Franz law:
\begin{equation}\label{eq2}
  \kappa_e=L\sigma T
\end{equation}
where L is the Lorenz number. So, there are similar dependencies of doping level and SOC between  $\mathrm{\kappa_e/\tau}$  and $\mathrm{\sigma/\tau}$.
The $\mathrm{\tau}$ is attained by comparing experimental electrical conductivity\cite{j4}of ZrTe with the calculated value of $\mathrm{\sigma/\tau}$ at room temperature, and the $\mathrm{\tau}$ is found to be  1.59 $\times$ $10^{-14}$ s.
 For  pristine
ZrTe,  the electronic thermal conductivity  along a(b) and c directions is 2.37 $\mathrm{W m^{-1} K^{-1}}$  and  2.90 $\mathrm{W m^{-1} K^{-1}}$, respectively.
 Using the calculated $\tau$, the electrical resistivity  of   pristine ZrTe along a(b) and c directions is 5.26$\times$$10^{-6}$ $\Omega$m and 3.66$\times$$10^{-6}$ $\Omega$m, respectively.
 A enhanced SOC effect on power factor  can be  observed along both a(b) and c directions, which is due to improved S caused by SOC.
 For  pristine ZrTe, the  power factor  along  a(b) and c directions is 3.82 $\mu$W/cm$\mathrm{K^2}$
  and 17.33 $\mu$W/cm$\mathrm{K^2}$, respectively, showing obvious anisotropy.

 Next, we obtain the scattering time  $\mathrm{\tau}$ at different temperatures by $\mathrm{\tau}$$\propto$$T^{-1}$\cite{j7}. For  pristine ZrTe, the  Seebeck coefficient S,  electrical resistivity $\mathrm{\rho}$ and  electronic thermal conductivity $\mathrm{\kappa_e}$  as a function of temperature with GGA+SOC are plotted \autoref{s-t}.
 It is found that the  Seebeck coefficient and  electrical resistivity along both a(b) and c directions firstly increase  with increasing temperature, and then decrease. It is noted that the Seebeck coefficient along c direction  changes from positive values to negative ones, when the temperature is larger than 870 K.
 At high temperature, the electrical resistivity  along both a(b) and c directions almost coincides. With increasing temperature,  the electronic thermal conductivity increases along both a(b) and c directions, showing weak  anisotropy.

\begin{figure}
   \includegraphics[width=8.0cm]{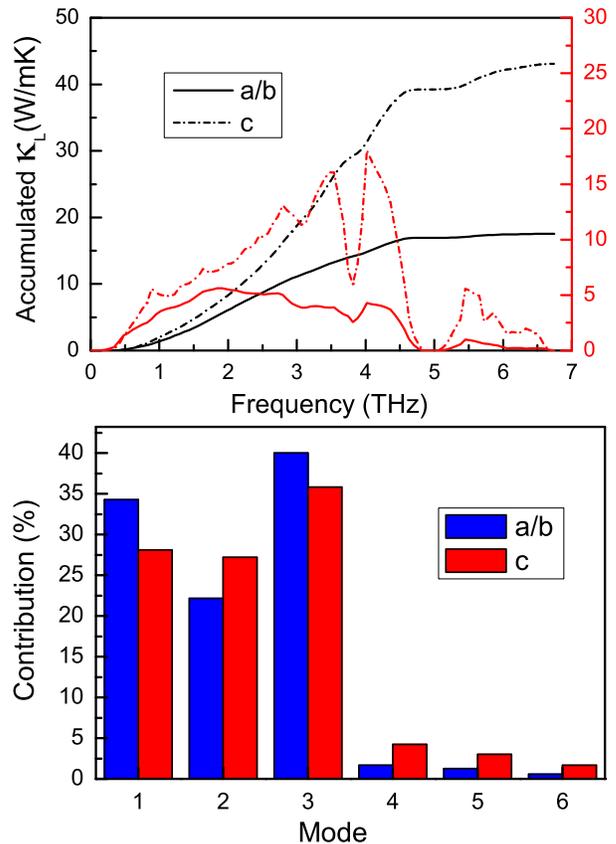}
  \caption{(Color online)At room temperature (300K), Top: the accumulated lattice thermal conductivities with respect to frequency, and the derivatives. Bottom: the phonon modes contributions along a(b) and c directions toward total lattice thermal conductivity. 1, 2, 3 represent TA1, TA2 and LA branches and 4, 5, 6 for optical branches.}\label{mkl}
\end{figure}
\begin{figure*}
  \includegraphics[width=16cm]{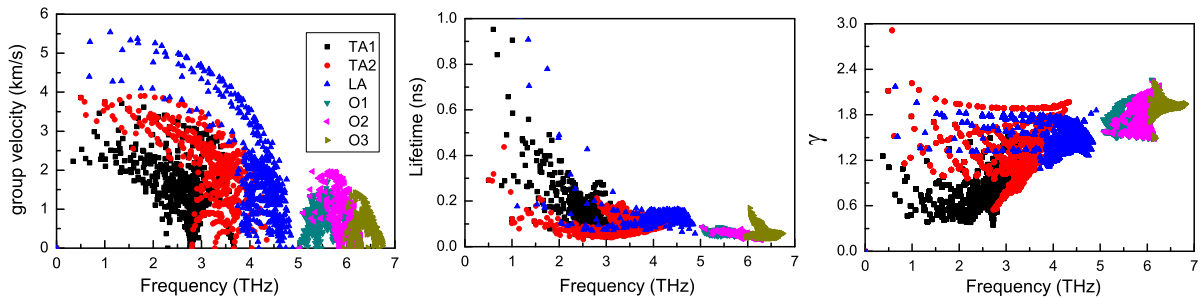}
  \caption{(Color online)The mode level phonon group velocities, phonon lifetimes (300K) and   Gr$\mathrm{\ddot{u}}$neisen parameters ($\gamma$) of pure ZrTe in the first BZ.}\label{v}
\end{figure*}

\subsection{PHONON TRANSPORT}
Based on the harmonic IFCs, the phonon dispersion of
 ZrTe can be attained along high-symmetry path, which along with atom partial density of states (DOS) are shown in \autoref{ph}.
 Due to each primitive cell containing  two atoms, there are  six vibrational branches consisting of
three acoustic and optical ones, respectively. No imaginary frequencies  in the phonon dispersion   indicate the thermodynamic stability of ZrTe.
The ZrTe belongs to  $P\bar{6}m2$ space group  whose  point group is  $D_{3h}$, and then
BZ-centre optical phonon modes of this crystal can be decomposed as
\begin{equation}\label{eq2}
  \Gamma=A_2+2E
\end{equation}
The $A_2$ and $E$ modes are  infrared-active, and $E$ mode is also  Raman-active.
The  phonon frequencies of $A_2$ and $E$  are shown in \autoref{tab1}.
It can be clearly seen that there is  a well-separated acoustic-optical gap of 0.15 THz  at the A point (0, 0, $\pi$/2) on the boundary of the BZ.
The top phonon band of the gap at the A point
is  doubly degenerate, while the bottom phonon band is a singlet state. Similar phonon gap can also be found in MoP\cite{j6}, which is larger than one of ZrTe.
However, the isoelectronic ZrSe shows no acoustic-optical gap, and  the presence of the
TDNPs of phonon has been predicted\cite{j8}.
From atom partial DOS, contribution to the  acoustic (optical) phonon branches  mainly comes from Te (Zr)  atoms.

\begin{table}[!htb]
\centering \caption{Theoretical optical phonon frequencies (THz) at the $\Gamma$ point; acoustic-optical gap (THz); lattice thermal conductance ($\mathrm{W m K^{-1}}$) }\label{tab1}
  \begin{tabular*}{0.48\textwidth}{@{\extracolsep{\fill}}cccccc}
  \hline\hline
$A_2$& $E$(1) &$E$(2) & $Gap$ &$\kappa_L(a/b)$&$\kappa_L(c)$\\\hline\hline
6.21& 6.07 &6.07&0.15&17.56&43.08\\\hline\hline
\end{tabular*}
\end{table}

 The intrinsic lattice thermal conductivity of ZrTe  can be attained from harmonic and anharmonic IFCs by solving the linearized phonon Boltzmann equation within single-mode RTA method. The phonon-isotope scattering is also considered, according to the formula proposed  by Shin-ichiro Tamura\cite{q24}.
The lattice thermal conductivities  of pure and   isotopic ZrTe along a(b) and c directions as a function of temperature are shown in \autoref{kl}.
In the considered temperature region, the intrinsic enhancement of phonon-phonon scattering with increasing temperature leads to
the decreased lattice thermal conductivity of ZrTe,  which typically results as $1/T$.
From \autoref{kl}, it is clearly seen  that the lattice thermal conductivity of ZrTe
 shows obvious anisotropy, where  the lattice thermal conductivity
along  c direction  is very  higher than that along  a(b)direction. Similar result can be found in MoP\cite{j6}, but is different from TaAs\cite{q12}.
At room temperature, the  lattice thermal conductivities of pure (isotopic) ZrTe  along a(b) and c directions are 17.56 (16.13)  $\mathrm{W m^{-1} K^{-1}}$ and  43.08 (37.82)  $\mathrm{W m^{-1} K^{-1}}$, which of pure ZrTe are shown in \autoref{tab1}.
To measure the anisotropic strength,  an anisotropy factor\cite{q12} is defined as $\eta=(\kappa_{L}(cc)-\kappa_{L}(aa))/\kappa_{L}(aa)$,
and the calculated  value is 145.3\%, which is larger than that of MoP, implying stronger anisotropy.
The lattice thermal conductivity is connected with Young's modulus by the simple relation $\kappa_L\sim \sqrt{E}$\cite{q16}. Calculated results show that the orders of Young's modulus and lattice thermal conductivity along a(b) and c directions are identical.
It is found that phonon-isotope scattering along c direction produces larger effects on lattice thermal conductivity than that along a(b) direction.
With  increasing temperature, isotopic effect on lattice thermal conductivity gradually decreases, which is due to improvement of phonon-phonon scattering.

At room temperature, the cumulative lattice thermal conductivities with respect to frequency along with the derivatives along a(b) and c directions are plotted in \autoref{mkl}. The cumulative thermal conductivity is defined by:
\begin{equation}\label{eq2}
  \kappa^c(\omega)=\int^\omega_0\Sigma_\lambda \kappa_\lambda\delta(\omega_\lambda-\omega^{'}) d \omega^{'}
\end{equation}
It is clearly seen  that the acoustic phonon branches dominate  lattice thermal conductivity, up to 96.49\% along a(b) direction and 91.12\% along c direction.
It is found that the optical contribution along c direction is larger than that along a(b) direction.
Furthermore, the relative contributions of six  phonon branches  to the total lattice
thermal conductivity along a(b) and c directions at 300K are plotted in \autoref{mkl}.
Along both a(b) and c directions, longitudinal acoustic (LA) phonon mode
 has larger contribution than any of  two transverse acoustic (TA1 or TA2) phonon modes. It is evident that optical
branches along c direction have obvious contribution.

The phonon transport  of ZrTe can be further understood  with the help of the mode level phonon group velocities
and lifetimes, which  are plotted in \autoref{v}.
The largest phonon group velocity  of  TA1, TA2 and LA branches in long-wavelength limit is 2.23 $\mathrm{km s^{-1}}$, 3.86 $\mathrm{km s^{-1}}$ and  4.40 $\mathrm{km s^{-1}}$, respectively.
 In the low frequency region,  the most of group
velocities of  TA2 and TA1 branches   are lower than those of LA branch, which leads to
main contribution to lattice thermal conductivity from LA branch.
 It is found that  the most of both group velocities and phonon lifetimes of acoustic  branches are larger than those of  optical branches, which  lead to the dominant contribution from  acoustic branches  to the total lattice thermal conductivity.
 The Gr$\mathrm{\ddot{u}}$neisen parameters ($\gamma$) can reflect the strength of anharmonic interactions, determining the intrinsic phonon-phonon scattering. Here, the Gr$\mathrm{\ddot{u}}$neisen parameters are  directly calculated by third order anharmonic IFCs.
The mode level  Gr$\mathrm{\ddot{u}}$neisen parameters of  pure ZrTe are shown in \autoref{v}.
The larger  $\gamma$ leads to lower  phonon lifetimes due to stronger anharmonicity, giving rise to lower lattice thermal conductivity.
It is clearly seen that mode phonon lifetimes and  Gr$\mathrm{\ddot{u}}$neisen parameters show the opposite trend.
For all  branches, the
$\gamma$ is fully positive.  The average  Gr$\mathrm{\ddot{u}}$neisen parameter is 1.52, indicating relatively strong anharmonic phonon scattering.
\begin{figure}
  \includegraphics[width=8cm]{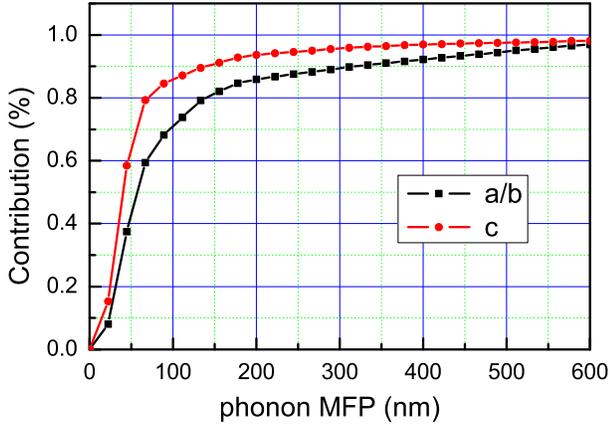}
  \caption{(Color online)For pure ZrTe, cumulative lattice thermal conductivity divided by total lattice thermal conductivity  with respect to phonon MFP along a/b and c directions at room temperature.}\label{mfp}
 \end{figure}

 To further understand the size dependence of lattice thermal
conductivity of ZrTe, the cumulative lattice thermal conductivity divided by total lattice thermal conductivity (CDT) with respect to MFP (300 K)  along a(b) and c directions are  plotted in \autoref{mfp}. The MFP cumulative lattice thermal conductivity is given by:
\begin{equation}\label{eq2}
  \kappa^c(l)=\int^l_0\Sigma_\lambda \kappa_\lambda\delta(l_\lambda-l^{'}) d l^{'}
\end{equation}
\begin{equation}\label{eq2}
  l_\lambda=|\mathrm{l_\lambda}|=|\mathrm{\nu_\lambda\otimes\tau_\lambda}|
\end{equation}
It can reflect the contribution to  total lattice thermal
conductivity from individual phonon modes with different MFP, namely it shows how phonons with
different MFP make contribution to the thermal conductivity.
It is clearly seen that the CDT along both a(b) and c directions approaches one with MFP increasing.
The contribution from phonons with MFP larger than 0.60 $\mathrm{\mu m}$ is very little.
 Phonons with MFP smaller than 0.13 (0.07) $\mathrm{\mu m}$ along a(b) direction and 0.07 (0.05) $\mathrm{\mu m}$ along c direction
 contribute  around  80\% (60\%)  to the lattice thermal conductivity.
 It is found that  phonons  dominating  the lattice thermal conductivity along a(b) direction  have  longer MFP than ones  along c direction.

 \begin{figure}
  \includegraphics[width=8cm]{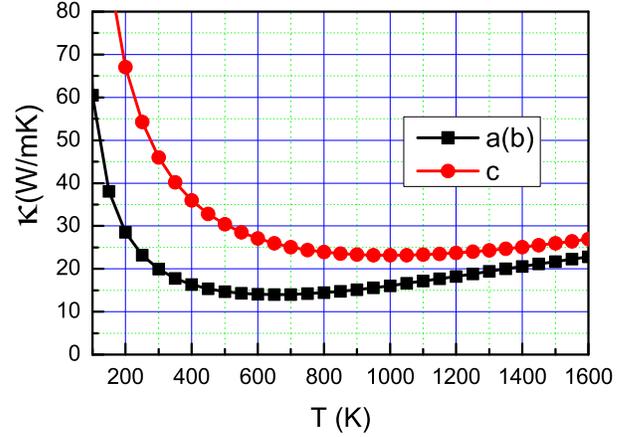}
  \caption{(Color online) The total  thermal conductivities  of  pristine ZrTe  as a function of temperature, including a(b)  and c directions. }\label{ktot}
\end{figure}

Based on calculated electronic and  lattice thermal conductivity,  the total  thermal conductivity can be attained, which is plotted in \autoref{ktot}. It is clearly seen that $\kappa$ firstly decreases with increasing temperature, and then increases.
It is because lattice part dominates thermal conductivity at low temperature, while electronic part is predominant at high temperature.
The minimum  $\kappa$ along a(b) and c directions is 13.98 $\mathrm{W m^{-1} K^{-1}}$  and 23.15  $\mathrm{W m^{-1} K^{-1}}$, respectively, and the
corresponding temperature is 650 K and 1000 K. The room-temperature $\kappa$ is 19.94 $\mathrm{W m^{-1} K^{-1}}$  and 45.98 $\mathrm{W m^{-1} K^{-1}}$, respectively. These results are useful for  the thermal management of ZrTe-based electronics devices.

\section{Conclusion}
The elastic properties and phonon transport  of the isostructural topological semimetal MoP has been investigated by the same method\cite{j6}.
It is found that the bulk modulus, shear modulus, Young's modulus of ZrTe are smaller than those of MoP, which means that ZrTe produces easily deformation
by applied external force.  The lattice thermal conductivity  of ZrTe (17.56  $\mathrm{W m^{-1} K^{-1}}$) along a(b) direction is very close to that of
MoP (18.41  $\mathrm{W m^{-1} K^{-1}}$), while the  lattice thermal conductivity  of ZrTe (43.08  $\mathrm{W m^{-1} K^{-1}}$) along c direction
is larger than that of MoP (34.71  $\mathrm{W m^{-1} K^{-1}}$).
It is noted that the average  lattice thermal conductivity ($\kappa_L(av)$=($\kappa_L(aa)$+$\kappa_L(bb)$+$\kappa_L(cc)$)/3) of ZrTe is slightly higher  than that of MoP.  This is because that ZrTe has larger phonon lifetimes and smaller Gr$\mathrm{\ddot{u}}$neisen parameters than MoP,  which  gives rise to higher lattice thermal conductivity for ZrTe than MoP.  Phonon transport of another classic topological  semimetal TaAs has been investigated\cite{q12}.
Any  of ZrTe, MoP and TaAs  shows  obviously  anisotropic  lattice thermal conductivity along a(b) and c directions. However, for TaAs, 
the lattice thermal conductivity  along a(b) direction  is larger than one along c direction, but the lattice thermal conductivity  of ZrTe or MoP along a(b) direction is smaller than that along c direction.

In summary,  the elastic and  transport properties of ZrTe are performed  by  the first-principles calculations and semiclassical Boltzmann transport theory.
The elastic tensor components $C_{ij}$ for ZrTe are presented,  which confirm the mechanical stability of the
structure. The bulk modulus, shear modulus, Young's modulus and Poisson's ratio are also attained by calculated  $C_{ij}$.
The  electronic  transport coefficients are also calculated within CSTA Boltzmann theory. For  pristine ZrTe, the  Seebeck coefficient,  electrical resistivity and  electronic thermal conductivity   will be of use for comparison with
future experimental measurements.
The  lattice thermal conductivity of ZrTe shows an obvious anisotropy along the a(b) and c crystal axis. It is found that isotope
scattering has observable effect on the lattice thermal conductivity, and  phonons with MFP larger than 0.60 $\mathrm{\mu m}$  have little contribution to the total lattice thermal conductivity.
The higher  lattice thermal conductivity of ZrTe than MoP can be explained by larger phonon lifetimes and smaller Gr$\mathrm{\ddot{u}}$neisen parameters. 
The total  thermal conductivity is also attained  for pristine ZrTe.
Our works shed light on elastic and transport properties of ZrTe, and will motivate farther experimental studies of elastic and  transport properties of  topological semimetals ZrTe.

\begin{acknowledgments}
This work is supported by the National Natural Science Foundation of China (Grant No.11404391). We are grateful to the Advanced Analysis and Computation Center of CUMT for the award of CPU hours to accomplish this work.
\end{acknowledgments}

\end{document}